\documentclass[final,10pt]{elsarticle}

\usepackage{amssymb}
\usepackage{amsmath}
\usepackage[dvipsnames]{xcolor}

\journal{Journal of Pure and Applied Algebra}
\let\OLDthebibliography\thebibliography
\renewcommand\thebibliography[1]{
  \OLDthebibliography{#1}
  \setlength{\parskip}{0pt}
  \setlength{\itemsep}{0pt plus 0.3ex}}
\begin{document}
\begin{frontmatter}

\title{Exact recursive calculation of circulant permanents: \\A band of different diagonals inside a uniform matrix}

\author[a]{Vitaly V. Kocharovsky\corref{*}}
\author[b]{Vladimir V. Kocharovsky}
\author[c]{\\Vladimir Yu. Martyanov}
\author[b]{Sergey V. Tarasov}

\cortext[*]{Corresponding author. \\{\it E-mail addresses}: 
vkochar@physics.tamu.edu (V.V. Kocharovsky),
kochar@appl.sci-nnov.ru (Vl.V. Kocharovsky), 
mavy7@mail.ru (V.Yu.Martyanov), 
serge.tar@gmail.com (S.V.Tarasov).}

\address[a]{Department of Physics and Astronomy, Texas A\&M University, College Station, TX 77843-4242, USA}

\address[b]{Institute of Applied Physics, Russian Academy of Science,
603950 Nizhny Novgorod, Russia}

\address[c] {Intel Corporation, 5000 W Chandler Blvd, Chandler, AZ 85226, USA}

\begin{abstract}
We present a finite-order system of recurrence relations for a permanent of circulant matrices containing a band of $k$ any-value diagonals on top of a uniform matrix (for $k = 1, 2$, and 3) as well as the method for deriving such recurrence relations which is based on the permanents of the matrices with defects. The proposed system of linear recurrence equations with variable coefficients provides a powerful tool for the analysis of the circulant permanents, their fast, linear-time computing and finding their asymptotics in a large-matrix-size limit. The latter problem is an open fundamental problem. Its solution would be tremendously important for a unified analysis of a wide range of the nature's $\sharp P$-hard problems, including problems in the physics of many-body systems, critical phenomena, quantum computing, quantum field theory, theory of chaos, fractals, theory of graphs, number theory, combinatorics, cryptography, etc. 
\end{abstract}

\end{frontmatter}

\section{Introduction. Significance and complexity of circulant permanents}

The permanent, $\textrm{per\,} C$, and the determinant, $\textrm{det\,} C$, of a $n \times n$ matrix $C$ correspond to two major operations --- the symmetrization and the anti-symmetrization, respectively. This fact predetermines their fundamental role in the quantum theory of many-body systems which are either bosonic (symmetric) or fermionic (anti-symmetric). The permanents are well-known in mathematical physics, especially, in quantum computing science and quantum field theory of interacting Bose fields \cite{Entropy2020,Caianiello1973,Wosiek1997,Scheel2004,Aaronson2011,Kalai2016,ShchesnovichPRL2016,Drummond2018}. However, compared to the determinants, the permanents are much more difficult to compute and they account for much more complicated many-body phenomena, such as the critical phenomena in phase transitions. For instance, an exact general solution of a long-standing three-dimensional Ising model \cite{Kadanoff} has been represented recently in terms of the permanent of a circulant matrix \cite{arXivIsing2016,PhysicaScripta2015,PLA2015Ising}.

The permanents have been studied in mathematics for more than a century (for a review, see \cite{Minc1978,Minc1987,Bapat2007,Brualdi2008,Stanley1-2012,Barvinok2016}), the most actively after discovery of the Ryser's algorithm \cite{Ryser}, publication of the comprehensive book "Permanents" \cite{Minc1978}, proof of the famous Valiant's theorem stating that their computing is a $\sharp P$-hard problem within the computational complexity theory \cite{Valiant1979}, and a recent development of a fully polynomial randomized approximation scheme \cite{Jerrum2004,Jerrum2015} for their computing. In fact, the permanents are intimately related to many fields of mathematics, including matrix algebra, combinatorics, number theory, theory of symmetric polynomials, discrete Fourier transform, $q$-analysis, computational complexity theory. For instance, in combinatorics matching in bipartite graphs is enumerated by permanents of 0-1 matrices \cite{Diaconis1999}. Another example is provided by the permanent of the Schur matrix which is equal to a sum that explicitly includes the ordinary M\"obius function of the number theory \cite{GrahamLehmer1976}. Importantly, permanents play a significant role in the algebraic complexity theory as universal polynomials \cite{Burgisser1997}. Nevertheless, despite an amazing recent activity and interest (see, e.g., \cite{Entropy2020,Aaronson2011,Kalai2016,ShchesnovichPRL2016,Drummond2018,Bapat2007,Barvinok2016,Jerrum2015,Gurvits2002,Gurvits2017,Fyodorov2006,Schwartz2009,FiedlerHall2012,Shchesnovich2016,Cifuentes2016,Glynn2010,Landsberg-JPAA2017} and references therein), the methods for analyzing and calculating the permanents and their asymptotics for most of large matrices remain illusive. 

The general deterministic algorithms \cite{Ryser,Glynn2010,Landsberg-JPAA2017} for computing the permanent of a $n \times n$ matrix with $n^2$ entries require exponentially large number of operations, at least $\sim n2^n$. A circulant matrix whose permanent has been analyzed by many authors \cite{Minc1978,King1969,Minc1987LAA,Codenotti1999,Thomas2004,LAA2017,Poi-JPAA2019} has just $n$ independent parameters as per Definition 2 below and requires smaller, but still exponentially large number of operations. The point is \cite{Codenotti2001} that a circulant graph on $n$ vertices with enough nonzero entries contains an arbitrary subgraph of size proportional to $n^{1/2}$. No efficient algorithm has been found for computing the permanent of a general circulant matrix of a large or moderate size in polynomial time. 

The situation could be very different for the permanent of matrices with a fixed number $k$ of any-value parameters which is not increasing with the matrix size $n$ or of a special structure, say, confining all nonzero entries to a band or restricted diagonal or rectangular blocks. One expects that for some classes of matrices with additional constraints computing their permanent is possible in polynomial time and, hence, accessible for applications. The latter has been demonstrated for matrices whose nonzero entries are confined to a band or just a few diagonals \cite{Minc1978,Schwartz2009,Poi-JPAA2019,Fonseca2010,Temme2012,Butera2015}. Efficient computing of the permanent of very sparse circulants \cite{Cifuentes2016,Codenotti2001,Servedio2005} is another example. In the present paper we consider an alternative class of matrices --- the circulant matrices composed of a band of $k$ any-value diagonals on top of a uniform $n \times n$ matrix.

The presence of nonzero, even constant entries spreading over an entire two-dimensional matrix plane in all directions drastically changes the behavior and asymptotics of the permanent and significantly complicates the problem of its calculation. The two aforementioned classes of matrices, with zero and nonzero entries outside a finite number of diagonals, constitute two different universality classes for permanents. Based on the reduction of the critical behavior in phase transitions to the asymptotics of the permanent of a mean-field correlation matrix \cite{arXivIsing2016,PhysicaScripta2015,PLA2015Ising}, these two classes could be related to two scaling asymptotics of the critical phenomena in the disordered and ordered phases, respectively. As is known from a phenomelogical renormalization-group approach \cite{Kadanoff,Goldenfeld}, they are very different since the correlation matrix abruptly tends to zero in the disordered phase, but remains nonzero for a very long distance due to long-range correlations in the ordered phase. Note that we can choose any (except the truly special, trivial zero) constant nonzero value, $c=1$ or $c \neq 1$, for the entries of the uniform background matrix (see Eq.~(\ref{Cck=1})) in the considered model of the circulant matrix with a band of $k$ any-value diagonals since it amounts to a simple rescaling of the permanent as is explained in Remark 1.3. Computing the permanent of such matrices is very important in a number of applications in mathematics, physics, computing, information systems, cryptography, etc. (see the references above). We aim at finding a system of recurrence relations for it. It would allow one to compute the permanent of such matrices in linear time. 

A base goal of the paper is to present a new method for finding the aforementioned system of recurrence relations: Recursion of permanents of the matrices with defects. A specific goal is to demonstrate practical ways of this method and communicate the new results on finding the system of recurrence relations for calculating nontrivial, multi-parametric circulant permanents with a band of $k = 1, 2$ or 3 different diagonals. In particular, such a recurrence, in the case of the zero-valued $k$-diagonal band, provides a solution to a famous problem of finding a finite recurrence for the $k$-m{\'e}nage numbers, or $k$-discordant permutations (for a discussion of $k$-m{\'e}nage numbers, see \cite{Riordan1958,Whitehead1979,Canfield1987,Bong1998,Flajolet2009,OEIS,Zeilberger2014,Alekseyev2016}).  

For calculating the permanent, especially of matrices whose entries are continuous variables, the proposed method of recursion of permanents with defects is more powerful and efficient than a widely known method of rook polynomials \cite{Riordan1958,Stanley1-2012,Whitehead1979,Canfield1987,Bong1998,Zeilberger2014}. The former incorporates an entire analytic, algebraic and combinatorial information on the permanental polynomial of $k$ variables. The latter is limited to just one artificial variable and mainly combinatorial information on the associated 0-1 matrices and is not directly suitable for finding the permanent as a function of all $k$ variables and revealing its asymptotics.   
 
{\bf Notation.} For brevity, we denote the permanent, $C_n = \textrm{per}\,C$, of any $n \times n$ matrix $C$ by a symbol of its size, $n$, placed next to the matrix's symbol as a subscript which simultaneously plays a part of a recursion variable.

\section{Definitions}

\noindent {\bf Definition 1.} The permanent and determinant of a $n\times n$ matrix $C=(c_p^q)$ are 
\begin{equation} 
C_n \equiv \textrm{per}\,C = \sum_{\sigma \in S_n} \prod_{p=1}^n c_p^{\sigma (p)} \quad \textrm{and} \quad
\textrm{det}\,C = \sum_{\sigma \in S_n} \textrm{sgn} (\sigma) \prod_{p=1}^n c_p^{\sigma (p)}.
\label{per-def}
\end{equation}
The sums run over symmetric group $S_n$, i.e., over permutations $\sigma$ of $1, 2,..., n$.

{\bf Definition 2.} The $n\times n$ circulant matrix $C=\textrm{Circ}(c_0, c_1 , ..., c_{n-1})$ with the $p$-th row and $q$-th column entries $c_p^q$ is the matrix with the rows obtained via consecutive cyclic permutations of the entries of the first row $(c_0, c_1 , ..., c_{n-1})$, or via the discrete Fourier transform of the set of its eigenvalues $\{ \lambda_l \mid l=1,...,n \}$:
\begin{equation}
c_p^q \equiv c_{q-p \ ({\rm mod} \ n)} = \frac{1}{n} \sum_{l=1}^n \lambda_l e^{2\pi i (q-p)(l-1)/n} , \qquad \lambda_l = \sum_{j=0}^{n-1} c_j e^{2\pi ij(l-1)/n} .
\label{circulant}
\end{equation} 
Thereinafter, we consider the circulant matrix $C=\textrm{Circ}(c_0,...,c_{k-1}, 1, ..., 1)$, say
\begin{equation}
C =\textrm{Circ}(c_0,c_1,c_2, 1, ..., 1)= \left[
    \begin{array}{ccccccccc}
     c_0 & c_1 & c_2 & 1 & \cdots  & 1 & 1 & 1 \\
     1 & c_0 & c_1 & c_2 & \cdots  & 1 & 1 & 1 \\
     1 & 1 & c_0 & c_1 & \cdots  & 1 & 1 & 1 \\
     1 & 1 & 1 & c_0 & \cdots  & 1 & 1 & 1 \\
     \vdots  & \vdots  & \vdots  & \vdots  & \ddots & \vdots  & \vdots  & \vdots  \\
     1 & 1 & 1 & 1 & \cdots  & c_0 & c_1 & c_2 \\
     c_2 & 1 & 1 & 1 & \cdots  & 1 & c_0 & c_1 \\
     c_1 & c_2 & 1 & 1 & \cdots  & 1 & 1 & c_0 \\
    \end{array}  \right] ,
    \label{C}
\end{equation} 
whose first row includes a set $\{c_0,..., c_{k-1}\}$ of just $k=1,2, \textrm{or} \ 3$ any-value entries while the rest of the first row's entries equal unity: $c_q = 1, q=k,...,n-1$. In other words, the circulant matrix $C$ contains a band of $k$ any-value diagonals inside a uniform $n \times n$ matrix $J=(J_p^q)$ with unity entries $J_p^q =1; \ p, q = 1, ..., n$.

It is convenient to derive the recurrence relations for its permanent on the basis of the following two matrices $A, B$ defined for any matrix size $n \geq k$.  
     
{\bf Definition 3.} Matrix $A = (a_p^q)$ is obtained from the original circulant $C=\textrm{Circ}(c_0,..., c_{k-1}, 1, ..., 1)$ by replacing all entries in the lower left ($p-q>n-k$) triangle of entries with unities (hereinafter, $\delta_{q,j}$ is the Kronecker delta),  
\begin{equation}
a_p^q = c_p^q \ \ \textrm{if} \ \ p-q \le n-k; \qquad a_p^q =1 \ \ \textrm{if} \ \ p-q>n-k .
\label{A}
\end{equation} 

{\bf Definition 4.} Matrix $B = (b_p^q)$ is obtained from a shifted circulant matrix 
\begin{equation}
\tilde{C}=\textrm{Circ}(c_1,..., c_{k-1}, 1, ..., 1, c_0) \equiv (\tilde{c}_p^q)
\label{tildeC}
\end{equation}
by replacing the upper rightmost and lower leftmost entries $\tilde{c}_1^n$ and $\tilde{c}_n^1$ with 1s,  
\begin{equation}
b_p^q = \tilde{c}_p^q + (1-\tilde c_n^1)\delta_{p,n}\delta_{q,1} + (1-\tilde c_1^n)\delta_{p,1}\delta_{q,n} .
\label{B}
\end{equation} 

{\bf Definition 5.} Two auxiliary matrices, $A' = ({a'_p}^q), R = (r_p^q)$, with entries 
\begin{equation}
{a'_p}^q = a_p^q +(c_2-1)\delta_{p,n}\delta_{q,1} , \qquad r_p^q = \tilde{c}_p^q + \sum_{l=2}^n (1-c_l^1)\delta_{p,l}\delta_{q,1} 
\label{A'R}
\end{equation} 
are used to prove the Theorem. The $A'$ differs from $A$, Eq.~(\ref{A}), just by the lowest left entry $c_2$. The $R$ is obtained from the shifted circulant $\tilde{C}$, Eq.~(\ref{tildeC}), by replacing all entries in the first column, except the upper one, with unity. 

{\bf Definition 6.} 
A symbol $M^{(l)}$ stands for a matrix $M = (m_p^q)$ whose $l$-th column is replaced with a column of 1s, that is, $m_p^{(l)q} = m_p^q + (1-m_p^q)\delta_{q,l}$. 

We found that in order to get the closed, finite-order recurrence relations we need to consider the permanent of the matrices with a unity-column defect, namely, $C^{(l)}_n, A^{(l)}_n, B^{(l)}_n$, as well as the following sums of such permanents

{\bf Definition 7} (where a sum index in parenthesis $(n)$ is a matrix size plus 1) 
\begin{equation}
A_{(n)} = \sum_{l=1}^{n-1} A^{(l)}_{n-1}, \quad B_{(n)} = \sum_{l=1}^{n-1} B^{(l)}_{n-1}, \quad C_{(n)} = \sum_{l=1}^{n-1} C^{(l)}_{n-1} = (n-1)C^{(1)}_{n-1} .
\label{sums}
\end{equation} 

{\bf Definition 8.} A star $(*)$-conjugation of any $k$-diagonal-band matrix $M$ with defects means a matrix $M(*)$ obtained from $M$ by just renaming its parameters in the inverse order: $c_0,...,c_{k-1} \rightarrow c_{k-1},..., c_0$. 

{\bf Definition 9.} $M(i|j)$ is the $(n-1) \times(n-1)$ submatrix obtained from a $n \times n$ matrix M after deleting the $i$-th row and $j$-th column. $M(i_1i_2|j_1j_2)$ is the $(n-2) \times(n-2)$ submatrix obtained from a $n \times n$ matrix M after deleting the $i_1$ and $i_2$ rows as well as the $j_1$ and $j_2$ columns.

\section{Method: Recursion of permanents with defects}     

We calculate the circulant permanent $C_n = \textrm{per}\,C$ via permanents of a few matrices obtained from the circulant $C$ by introducing certain defects, like those in the definitions 3--9. The set of those matrices with defects is dictated by a requirement of its self-closure in the course of the reducing their permanents to the permanents of lower-size matrices by means of the Laplace expansion \cite{Minc1987}, certain relations between different permanents valid due to particular patterns and symmetries of the $k$-diagonals-band circulants, and various matrix transformations leaving the permanent invariant, such as transposing with respect to the main or minor diagonals, permutations of rows or/and columns, etc. In particular, the matrices like $A$ and $B$ in Eqs.~(\ref{A}) and (\ref{B}) have the non-unity entries just within the main band of the $k$ diagonals. The matrix $R$ in Eq.~(\ref{A'R}) whose first-column entries are all unities, except the upper one, is instrumental for establishing a recursive coupling with the sums of permanents, Eq.~(\ref{sums}), via the Laplace expansion over the first row. Some important matrices with defects arise naturally as minors in the Laplace expansion or Lemmas 1 and 2 below. 

In general, we introduce the matrices with defects located as close to the corners of the matrices as possible in order to avoid recursive chains of matrices with the defects penetrating deeper and deeper inside the matrices. 

The permanent of each matrix with defects is associated with a particular combinatorial meaning. Say, the permanents of the circulant $C$ and matrix with defects $A$, in the case of zero parameters, $c_0=...=c_{k-1}$, are equal to the famous $k$-m{\'e}nage numbers for circular- and straight-table problems, respectively. 

One of the key ingredients of the method is an explicit involvement of the sums of permanents over the matrices with the unity-column or unity-row defects (see Eq.~(\ref{sums})) into the permanent's recurrence relations. 

The star $(*)$-conjugation introduced in Definition 8 reveals an important symmetry and helps to reduce the number of matrices with defects needed for achieving completeness of the system of recurrence relations.

Especially, we employ the following lemmas which immediately follow from the permanent's definition and the Laplace expansion of the permanent along the $p^*$-th row or the $p^*$-th row and $q^*$-th column, etc. 

{\bf Lemma 1.} If a matrix $\tilde{M}$ differs from a matrix $M = (m_p^q)$ by just the entries (defects) $\tilde{m}_{p^*}^q$ in a single row $p=p^*$, then its permanent $\tilde{M}_n = \textrm{per} \,\tilde{M}$ differs from the permanent $M_n = \textrm{per}\,M$ of the unperturbed matrix by just the sum over separate linear corrections per each $p^*$-row entry, $(\tilde{m}_{p^*}^q - m_{p^*}^q)$, multiplied by the corresponding permanental minor, i.e., by the permanent $M(p^*|q)_{n-1} = \textrm{per}\,M(p^*|q)$ of the submatrix $M(p^*|q)$ of the lower size $n-1$, 
\begin{equation}
\tilde{M}_n = M_n + \sum_{q=1}^n (\tilde{m}_{p^*}^q - m_{p^*}^q) M(p^*|q)_{n-1} . 
\label{row-defects}
\end{equation} 
A similar representation is valid when all defects are located in a single column.

{\bf Lemma 2.} If a matrix $\tilde{M}$ differs from a matrix $M = (m_p^q)$ by just the entries (defects) $\tilde{m}_{p^*}^q$ in a row $p=p^*$ and $\tilde{m}_p^{q^*}$ in a column $q=q^*$, then its permanent $\tilde{M}_n = \textrm{per}\,\tilde{M}$ differs from the permanent $M_n = \textrm{per}\,M$ of the unperturbed matrix by the following superposition of (i) the separate linear corrections per each $p^*$-row entry and each $q^*$-column entry, multiplied by a corresponding permanental minor of the size $n-1$, and (ii) the cross-correlated quadratic corrections per each pair of defects, one from $p^*$-row and one from $q^*$-column, multiplied by the permanent of the corresponding submatrix $M(p^*p|qq^*)$ of the size $n-2$, 
\begin{multline}
\tilde{M}_n = M_n + \sum_{q=1}^n (\tilde{m}_{p^*}^q - m_{p^*}^q) M(p^*|q)_{n-1} + \sum_{p=1, p\neq p^*}^n (\tilde{m}_{p}^{q^* }- m_{p}^{q^*}) M(p|q^*)_{n-1}
\\
+ \sum_{q=1, q\neq q^*}^n \ \sum_{p=1, p\neq p^*}^n (\tilde{m}_{p^*}^q - m_{p^*}^q) (\tilde{m}_{p}^{q^* }- m_{p}^{q^*}) M(p^*p|qq^*)_{n-2} \ .
\label{row+column-defects}
\end{multline} 
Note that the defect $\tilde{m}_{p^*}^{q^*}$ at the intersection of the $p^*$-th row and $q^*$-th column contributes to the right hand side of Eq.~(\ref{row+column-defects}) just once through a linear correction and does not contribute at all to the cross-correlated quadratic corrections.  

A generalization of Lemmas 1 and 2 to a general case when the defects are located in the three or more different rows and columns is straightforward.

\section{The permanent of a uniform circulant matrix with one any-value diagonal (k=1) and the rencontres numbers}

In this section we consider the simplest nontrivial case of a uniform circulant $n \times n$ matrix with a band of $k$ any-valued diagonals, namely, the case of just one $(k=1)$ diagonal with the entries $c_0$ inside the matrix $J$ of all 1s: 
\begin{equation}
C = \left[
    \begin{array}{ccccccccc}
     c_0 & 1 & \cdots  & 1 & 1 \\
     1 & c_0 & \cdots  & 1 & 1 \\
     \vdots  & \vdots  & \ddots & \vdots  & \vdots  \\
     1 & 1 & \cdots  & c_0 & 1 \\
     1 & 1 & \cdots  & 1 & c_0 \\
    \end{array}
    \right] \quad .
    \label{Ck=1}
    \end{equation}
{\bf Proposition 1.} The recurrence equation for the permanent of this circulant is  
\begin{equation}
C_n = nC_{n-1} +(c_0-1) [C_{n-1} - (n-1)C_{n-2}], \ \ \textrm{or} \ \ C_n - nC_{n-1} = (c_0-1)^n ,   
    \label{REk=1}
\end{equation}
and yields the exact solution for the permanent, 
\begin{equation}
C_n = e^{c_0-1} \Gamma(n+1,c_0-1) ,
    \label{C_nk=1}
\end{equation}
via the well-known upper incomplete gamma function 
\begin{equation}
\Gamma(n+1,x) = \int_x^{\infty} t^n e^{-t} dt . 
    \label{GammaF}
\end{equation}
{\bf Proof.} The permanent is given by the Laplace expansion over the first row as
\begin{equation}
C_n = c_0C_{n-1} + C_{(n)} 
    \label{C_nViaC_(n)}
\end{equation}
via the sum of permanents $C_{(n)}$, Eq.(\ref{sums}), which consists of $n-1$ equal permanents of the circulant matrices with a 1's-column defect $C^{(q)}_{n-1} = C^{(1)}_{n-1} (q=1,...,n-1)$. The latter can be evaluated by means of Lemma 1, and so we have
\begin{equation}
C_{(n)} = (n-1)C^{(1)}_{n-1} , \qquad C^{(1)}_{n-1} = C_{n-1} + (1-c_0)C_{n-2} .
    \label{C^(1)}
\end{equation}
The first part of Eq.~(\ref{REk=1}) follows from (\ref{C_nViaC_(n)}), (\ref{C^(1)}). It is immediate to solve it for 
\begin{equation}
C_n - nC_{n-1} = (c_0-1)^n  
    \label{1stOrderREk=1}
\end{equation}
via a geometrical progression that could be started from the matrix size $n=2$, for which the relevant permanents are easy to calculate from the definition (\ref{per-def}) ($C_1=c_0, C_2=c_0^2+1$). It is valid even for $n=1$ if one assigns a unity value for the permanent of the zero-size matrix, $C_0=1$, so that $C_1-C_0=c_0-1$. The latter Eq.~(\ref{1stOrderREk=1}) is exactly the known recurrence equation for the upper incomplete gamma function which follows from integration by parts of Eq.~(\ref{GammaF}). This fact proves Eq.~(\ref{C_nk=1}) and completes the proof of the Proposition 1.

{\bf Remark 1.1.} According to the definition (\ref{per-def}), the permanent of the matrix (\ref{Ck=1}) is a polynomial of one variable $c_0$ of the order $n$. It can be represented as 
\begin{equation}
C_n = \sum_{j=0}^n P_j c_0^j = n! \sum_{j=0}^n (c_0-1)^j/j! \quad .
    \label{polynomialCk=1}
\end{equation}
Here the second representation follows from Eq.~(\ref{C_nk=1}) and the known fact that a Taylor series of the function $e^x\Gamma(n+1,x) = n! \sum_{j=0}^n x^j/j!$ becomes finite at the integer values of $n = 0, 1, 2,...$. Thus, the analytic formula in Eq.~(\ref{C_nk=1}) gives an exact solution for this permanent and, importantly, allows one to find its asymptotics in the limit of a large matrix size $n \to \infty$ via the known properties of the upper incomplete gamma function. 

{\bf Remark 1.2.} Famous derangement numbers, or subfactorials, \cite{Stanley1-2012} counting the number of permutations of the set $\{ 1,...,n\}$ that have no fixed points, 
\begin{equation}
D_{n,0} \equiv \ !n = \lfloor n!/e \rceil = n!\sum_{j=0}^n (-1)^j/j! = \int_0^{\infty} (t-1)^ne^{-t}dt,
\label{derangements}
\end{equation}
and more general rencontres numbers counting the number of permutations of the set $\{ 1,...,n\}$ with exactly $k$ fixed points ($0\le k \le n$),
\begin{equation}
D_{n,k} = \frac{n!}{k!(n-k)!} D_{n - k,0} \quad ,
\label{recontres}
\end{equation}
are closely related to the permanent of the matrix (\ref{Ck=1}). Namely, the number of derangements is equal to the permanent of the matrix $C$ with the zero diagonal, that is, just to the constant term $P_0$ of the permanental polynomial in Eq.~(\ref{polynomialCk=1}),
\begin{equation}
D_{n,0} = C_n|_{c_0=0} = P_0 \ .
\label{recontres=P0}
\end{equation}

{\bf Remark 1.3.} The permanent $C'_n$ of a somewhat more general circulant $n \times n$ matrix $C'$ with one $c_0$-diagonal inside a uniform matrix with one and the same value $c$, not necessarily unity, assigned to all other entries,
\begin{equation}
C' = \left[
\begin{array}{ccccccccc}
     c_0 & c & \cdots  & c & c \\
     c & c_0 & \cdots  & c & c \\
     \vdots  & \vdots  & \ddots & \vdots  & \vdots  \\
     c & c & \cdots  & c_0 & c \\
     c & c & \cdots  & c & c_0 \\
    \end{array}
    \right] \quad ,
    \label{Cck=1}
\end{equation}
can be easily reduced to the one in Eq.~(\ref{C_nk=1}) by means of rescaling:
\begin{equation}
C'_n = c^n e^{\frac{c_0}{c}-1} \Gamma(n+1,\frac{c_0}{c}-1) .
    \label{C'_nk=1}
\end{equation}
This formula as well as its analogs for the uniform circulant matrices with a band of $k>1$ diagonals are useful for analyzing an overall scaling of the permanents. Yet, thereinafter we'll skip such a straightforward generalization and set $c=1$.

{\bf Remark 1.4.} Another derivation of Eq.~(\ref{C'_nk=1}) is based on the combinatorics of the permanent in the definition (\ref{per-def}) and the rencontres numbers in Eq.~(\ref{recontres}):
\begin{equation}
C'_n = \sum_{j=0}^n D_{n,j} c^j c_0^{n-j} = \int_0^{\infty} \frac{(c-c_0+ct)^n}{e^t}dt = c^n e^{\frac{c_0}{c}-1} \Gamma(n+1,\frac{c_0}{c}-1) .
    \label{recontresC'_nk=1}
\end{equation}

{\bf Remark 1.5.} In the case $k=0$ of the matrix which has all unity entries without a band of any-value diagonals, $C=J$, it is easy to find the permanent directly from the definition in Eq.~(\ref{per-def}) by means of a trivial combinatorics: $\textrm{per}\,J =n!$. Of course, the same result follows from the general analytic formula in Eq.~(\ref{C_nk=1}) in the corresponding particular case $c_0=1$.

\section{The permanent of a uniform circulant matrix with a band of two any-value diagonals (k=2) and the m{\'e}nage numbers}

Here we consider a more envolved, but still simple case of a uniform circulant $n \times n$ matrix with a band of two $(k=2)$ diagonals (with the entry values $c_0$ and $c_1$) inside the matrix $J$ of all 1s: 
\begin{equation}
C =\textrm{Circ}(c_0,c_1,1, ..., 1)= \left[
    \begin{array}{ccccccccc}
     c_0 & c_1 & 1 & \cdots  & 1 & 1 & 1 \\
     1 & c_0 & c_1 & \cdots  & 1 & 1 & 1 \\
     1 & 1 & c_0 & \cdots  & 1 & 1 & 1 \\
     \vdots  & \vdots  & \vdots  & \ddots & \vdots  & \vdots  & \vdots  \\
     1 & 1 & 1 & \cdots  & c_0 & c_1 & 1 \\
     1 & 1 & 1 & \cdots  & 1 & c_0 & c_1 \\
     c_1 & 1 & 1 & \cdots  & 1 & 1 & c_0 \\
    \end{array}  \right] .
    \label{Ck=2}
    \end{equation}
{\bf Proposition 2.} The permanent of this circulant with a band of two diagonals,
\begin{equation}
C_n \equiv \textrm{per Circ}(c_0,c_1,1,...,1) = A_n+(c_1-1)[A_{n-1} +(c_1-c_0)A^{(1)}_{n-1} +A_{(n)} -A(*)_{(n)}],
    \label{C_nk=2}
\end{equation}
is given by a solution of the following system of recurrence relations   
\begin{equation}
\begin{split}
&A_n = c_0A_{n-1} +(n-2+c_1)A^{(1)}_{n-1} +(1-c_1)A(*)_{(n-1)} ,
\\
&A^{(1)}_n = A_{n-1} +(n-2+c_1)A^{(1)}_{n-1} +(1-c_1)A(*)_{(n-1)} ,
\\
&A_{(n)} = (n-1)A^{(1)}_{n-1} +(1-c_1)A(*)_{(n-1)} ,
\\
&A(*)_{(n)} = (n-1)A^{(1)}_{n-1} +(1-c_0)A_{(n-1)} 
\label{REk=2}
\end{split}
\end{equation}
with the initial conditions set at the starting recursive order $n=1$ as follows
\begin{equation}
A_1 = c_0, \quad A^{(1)}_1 = 1, \quad A_{(1)} = A(*)_{(1)} = 0 .
    \label{ICk=2}
\end{equation}
{\bf Proof.} The permanent $C_n$ is given by Lemma 1 applied to the matrix $A$, Eq.~(\ref{A}), which differs from the circulant $C$ by one defect in the lower left corner:
\begin{equation}
C_n = A_n +(c_1-1)A(*)_{n-1} .
    \label{L1Ck=2}
\end{equation}
Here the $(*)$-conjugation means renaming $c_0 \to c_1, c_1 \to c_0$ as per Definition 8. The permanent $A_n$ is given by the Laplace expansion over the first row,
\begin{equation}
A_n = c_0A_{n-1} +(c_1-1)A^{(1)}_{n-1} +A_{(n)} ,
    \label{A_nk=2}
\end{equation}
via the sum $A_{(n)}$, Eq.~(\ref{sums}), which we find via summation of the analogs of Eq.~(\ref{L1Ck=2}) written for the circulant with a 1's-column defect at the position $q=l$,
\begin{equation}
C^{(l)}_n = A^{(l)}_n +(c_1-1)(1-\delta_{l,1})A^{(n+1-l)}(*)_{n-1} , \quad l=1,2,...,n .
    \label{L1C^(l)k=2}
\end{equation}
Taking into account that the permanent of the circulant matrix with a 1's-column defect does not depend on the position $l$ of the 1's-column defect and, hence, equals to $C^{(l)}_n = A^{(1)}_n \ \forall l$, we immediately get the third equation of the system (\ref{REk=2}). The forth equation is just its $(*)$-conjugated counterpart. Plugging in the third equation into Eq.~(\ref{A_nk=2}), we get the first equation of the system (\ref{REk=2}). At last, the second equation of the system (\ref{REk=2}) follows from the first equation and the representation of the permanent of the matrix $A^{(1)}$ with a 1's-column defect at the position of the first column obtained by means of Lemma 1,
\begin{equation}
A^{(1)}_n = A_n +(1-c_0)A_{n-1} .
    \label{L1A^(1)k=2}
\end{equation}

Eq.~(\ref{C_nk=2}) for the circulant permanent $C_n$ itself follows from Eq.~(\ref{L1Ck=2}) after plugging in the equation for the $(*)$-conjugated permanent of the matrix $A$,
\begin{equation}
A(*)_{n-1} = A_{n-1} + (c_1-c_0)A^{(1)}_{n-1} + A_{(n)} -A(*)_{(n)} .
    \label{A(*)_nk=2}
\end{equation}
The latter equation is a consequence of the identity $A^{(1)}_n = A^{(1)}(*)_n$ with both sides substituted with an explicit expression
\begin{equation}
A^{(1)}_n = A_{n-1} +(c_1-1)A^{(1)}_{n-1} +A_{(n)} 
    \label{A^(1)viaAnA_(n)k=2}
\end{equation}
which follows from Eqs.~(\ref{L1A^(1)k=2}) and (\ref{A_nk=2}). The initial conditions (\ref{ICk=2}) provide the correct values for all permanents at the recursive order $n=2$, namely,
\begin{equation}
A_2 = c^2_0 +c_1, \quad A^{(1)}_2 = c_0 +c_1, \quad A_{(2)} = A(*)_{(2)} = 1, \quad A(*)_2 = c^2_1 +c_0.
    \label{ICn=2k=2}
\end{equation}
This completes the proof of the Proposition 2.

{\bf Remark 2.1.} The permanent of the uniform circulant matrix $C$ with a band of two diagonals ($k=2$), Eq.~(\ref{Ck=2}), is given by the system of linear homogeneous recurrence relations with variable coefficients of the 4-th order, Eq.~(\ref{REk=2}), whereas in the case of a band of one diagonal ($k=1$) the system of recurrence relations is of the 2-nd order, Eq.~(\ref{REk=1}). The system in Eq.~(\ref{REk=2}) allows one to compute the entire permanental polynomial, 
\begin{equation}
\textrm{per}\,C = \sum_{j_0,j_1=0}^n P_{j_0,j_1} \ c_0^{j_0} c_1^{j_1} ,
\label{polynomialk=2}
\end{equation}
as a function of two valiables ($c_0, c_1$) for any large matrix size $n$ and to find its asymptotics analytically. The relevant results will be presented elsewhere. 

{\bf Remark 2.2.} Famous m{\'e}nage numbers \cite{Muir1882,KaplanskyRiordan1946,Touchard1953,Whitehead1979,Canfield1987,Flajolet2009,OEIS,Zeilberger2014} $U_n$ counting the number of different ways to seat $n$ husbands at a circular table of $2n$ places so that men and women alternate and no adjacent couples are allowed, that is, the number of 2-discordant permutations $\sigma$ of $\{ 1,...,n\}$ such that $\sigma(j)$ is not congruent to any $j, j+1 \ (\textrm{mod} \, n)$, are equal to the permanent of the uniform circulant matrix $C$ with a band of two zero diagonals. That is, they are equal to a particular value of the permanent when both variables are zero, $c_0=c_1=0$, which is the constant term of the permanental polynomial in Eq.~(\ref{polynomialk=2}), 
\begin{equation}
U_n = C_n|_{c_0=c_1=0} = P_{0,0} \ .
\label{menage=P00}
\end{equation}
A well-known fourth-order recurrence equation for the m{\'e}nage numbers \cite{Muir1882,Canfield1987},
\begin{equation}
U_n = nU_{n-1} +2U_{n-2} -(n-4)U_{n-3} - U_{n-4} ,
\label{REmenage}
\end{equation}
immediately follows from the recurrence relations for the permanent $C_n$, Eq.~(\ref{REk=2}), as its particular case. Namely, for $c_0=c_1=0$ the permanent and, hence, the m{\'e}nage numbers (the sequence A000179 in \cite{OEIS}) are given by a simple formula 
\begin{equation}
U_n = C_n = A_n -A_{n-1} 
\label{menageC_n}
\end{equation}
via the permanent $A_n$ of the matrix with the defect $A$, Eq.~(\ref{A}), which coincides with the straight m{\'e}nage number $V_n$, counting the number of permutations $\sigma$ of $\{ 1,...,n\}$ such that $\sigma(j)$ is not congruent to any $j, \textrm{min} \{j+1,n\} \ (\textrm{mod} \, n)$ (that corresponds to an analogous problem of seating $n$ husbands at a straight-line table), and satisfies the third-order recurrence (the sequence A000271 in \cite{OEIS})
\begin{equation}
A_n = (n-1)(A_{n-1} +A_{n-2}) +A_{n-3}.  
\label{REmenageA_n}
\end{equation}
The latter immediately follows from Eq.~(\ref{REk=2}) since 
\begin{equation}
A^{(1)}_n = A_n +A_{n-1} \quad \& \quad A_{(n)} = A(*)_{(n)} = A_n +A_{n-1} +A_{n-2} \quad \textrm{for} \ c_0=c_1=0.
\label{menageA^(1)A_(n)}
\end{equation}
So, the method of recursion of permanents with defects provides a simple derivation of the recurrence (\ref{REmenageA_n}) for the straight m{\'e}nage numbers $V_n = A_n|_{c_0=c_1=0}$ which is different from a long combinatorial derivation (see Theorem 1 in \cite{Canfield1987}).

\section{The permanent of a uniform circulant matrix with a band of three any-value diagonals (k=3) and the $3$-m{\'e}nage numbers}

\noindent Finally, we consider a quite complicated case of a uniform circulant $n \times n$ matrix with a band of three $(k=3)$ diagonals (with the entry values $c_0, c_1$, $c_2$) inside the matrix $J$ of all 1s. It is depicted as $C$ in Eq.~(\ref{C}) or $\tilde{C}$, Eq.~(\ref{tildeC}), below
\begin{equation}
\tilde{C}=\textrm{Circ}(c_1,c_2, 1, ..., 1, c_0) = \left[
    \begin{array}{ccccccccc}
     c_1 & c_2 & 1 & \cdots  & 1 & 1 & c_0 \\
     c_0 & c_1 & c_2 & \cdots  & 1 & 1 & 1 \\
     1 & c_0 & c_1 & \cdots  & 1 & 1 & 1 \\
     \vdots  & \vdots  & \vdots  & \ddots & \vdots  & \vdots  & \vdots  \\
     1 & 1 & 1 & \cdots  & c_1 & c_2 & 1 \\
     1 & 1 & 1 & \cdots  & c_0 & c_1 & c_2 \\
     c_2 & 1 & 1 & \cdots  & 1 & c_0 & c_1 \\
    \end{array}  \right] . 
\label{matrixtildeC}
\end{equation}

\noindent \hspace{2mm}{\bf Theorem.} The permanent of this circulant with a band of three diagonals is
\begin{multline}
C_n \equiv \textrm{per Circ}(c_0,c_1,c_2,1,...,1) = B_n +(c_0-1)A_{n-1} +(c_2-1)A(*)_{n-1} 
\\
+ (c_0 - 1) (c_2 - 1) B_{n - 2}
\label{C_nk=3}
\end{multline}
and is determined by a solution of the following system of recurrence relations:
\begin{equation}
A_n = A_{(n)} -(1-c_2)A^{(2)}_{n-1} +(c_0+c_1-1)A_{n-1} -(1-c_0)(1-c_1)A_{n-2},
    \label{REA}
\end{equation}
\begin{multline}
B_n = B_{(n)} +c_1B_{n-1} -(1-c_0)A_{n-1} -(1-c_2)A(*)_{n-1}
\\
-(1-c_0)^2A_{n-2} -(1-c_2)^2A(*)_{n-2} +(1-c_0)(1-c_2)B_{n-2} ,
\label{REB}
\end{multline}
\begin{multline}
A_{(n)} = (n-3)C^{(1)}_{n-1} +A^{(2)}_{n-1} +A_{n-1} +2(1-c_2)A(*)_{(n-1)} +(1-c_1)B_{(n-1)}
\\
+(1-c_0)A_{n-2} +(c_1+c_2-2)A(*)_{n-2} -c_2(1-c_2)A(*)_{(n-2)} -2(1-c_0)(1-c_2)B_{(n-2)}
\\
-(2+c_0-c_1-c_2)(1-c_2)A(*)_{n-3} +(1-c_0)(1-c_2)^2 A(*)_{n-4} ,
\label{REA_(n)}
\end{multline}
\begin{multline}
B_{(n)} = (n-3)C^{(1)}_{n-1} +A_{n-1} +A(*)_{n-1} +(1-c_0)A_{(n-1)} +(1-c_2)A(*)_{(n-1)}
\\
-(1-c_0)(1-c_2)B_{(n-2)} -(1-c_0)^2 A_{n-3} -(1-c_2)^2 A(*)_{n-3} ,
\label{REB_(n)}
\end{multline}
\begin{multline}
C^{(1)}_n = (n-3)C^{(1)}_{n-1} +A_{n-1} +A(*)_{n-1} +B_{n-1} 
\\
+(1-c_0)[A_{(n-1)} -A_{n-1}] +(1-c_2)[A(*)_{(n-1)} -A(*)_{n-1}] -(1-c_0)(1-c_2)B_{(n-2)} 
\\
-(1-c_0)^2 [A_{n-2} +A_{n-3}] -(1-c_2)^2 [A(*)_{n-2} +A(*)_{n-3}] .
\label{REC}
\end{multline} 

The symbols $A_{(n)}$, $B_{(n)}$ in Eqs.~(\ref{REA}), (\ref{REB}) stand for the right hand sides of Eqs.~(\ref{REA_(n)}), (\ref{REB_(n)}), respectively. In Eqs.~(\ref{REA}) and (\ref{REA_(n)}) the symbol $A^{(2)}_{n-1}$ stands for
\begin{multline}
A^{(2)}_{n-1} = C^{(1)}_{n-1} +(1-c_1)A(*)_{n-2} +(1-c_2)A(*)_{(n-2)} 
\\
+ (1+c_0-c_1)(1-c_2)A(*)_{n-3} -(1-c_0)(1-c_2)^2 A(*)_{n-4} .
    \label{REA^(2)}
\end{multline}

{\bf Remark 3.1.} The recurrence system includes the $(*)$-conjugated counterparts of Eqs.~(\ref{REA}) and (\ref{REA_(n)}) for $A(*)_n$ and $A(*)_{(n)}$. Their explicit form is given in the Appendix. Obviously, $B(*)_n \equiv B_n$, $B(*)_{(n)} \equiv B_{(n)}$ and $C(*)^{(1)}_n \equiv C^{(1)}_n$. Hence, the $(*)$-conjugated counterparts of Eqs.~(\ref{REB}), (\ref{REB_(n)}), (\ref{REC}) are not needed. Note that in the present case of $k=3$ the star $(*)$-conjugation means just renaming $c_0 \to c_2, c_2 \to c_0$ and leaves $c_1$ untouched as per Definition 8.

{\bf Remark 3.2.} The system consists of the three blocks of different origin: 

\noindent(i) Block 1 --- the basis permanents (\ref{REA}), (\ref{REB}) of the matrices with entry defects; 
\noindent(ii) Block 2 --- the sums of permanents (\ref{REA_(n)}), (\ref{REB_(n)}) with a 1's column defect; 

\noindent(iii) Block 3 --- the special permanent (\ref{REC}) of the matrix $C$ with all 1s in the first column. 

{\bf Proof of the Theorem.} While applying Lemmas 1 and 2 below, we'll employ the auxiliary matrices $A'$ and $R$ defined in Eq.~(\ref{A'R}). They differ from the basis matrices $A$ and $B$ by one and two defects, so that their permanents are given by Lemma 1 and Lemma 2, respectively, as follows
\begin{equation}
A'_n = A_n + (c_2-1)B_{n-1} ,
    \label{A'_n}
\end{equation}
\begin{equation}
R_n = B_n - (1-c_0)(1-c_2)B_{n-2} .
    \label{R_n}
\end{equation}
Eq.~(\ref{C_nk=3}) follows from Lemma 2 applied to $B$ which differs from ${\tilde C}$ by two defects.

Eqs.~(\ref{REA}) and (\ref{REB}) follow from the Laplace expansions over the first row:
\begin{equation}
A_n = c_0A_{n-1} + (c_1-1)A^{(1)}_{n-1} +(c_2-1)A^{(2)}_{n-1} +A_{(n)} ,
    \label{LaplaceA_n}
\end{equation}
\begin{equation}
R_n = c_1B_{n-1} + (c_2-1)B^{(1)}_{n-1} +(c_0-1)B^{(n-1)}_{n-1} +B_{(n)} .
    \label{LaplaceR_n}
\end{equation}

Eq.~(\ref{REA_(n)}) is based on Lemma 2 applied to $A$ viewed as $C$ with three defects:
\begin{equation}
A_n = C_n + (1-c_1)B_{n-1} +2(1-c_2)A'(*)_{n-1} +(1-c_2)^2 A(*)_{n-2} .
    \label{A_nViaC_n}
\end{equation}
Namely, we use an analog of Eq.~(\ref{A_nViaC_n}) for a matrix $A^{(q)}$ with a 1's-column defect:
\begin{multline}
A^{(q)}_n = C^{(q)}_n + (1-c_1)B^{(q-1)}_{n-1}(1-\delta_{q,1}) +(1-c_2)^2 A^{(n+1-q)}(*)_{n-2}(1-\delta_{q,1})(1-\delta_{q,2}) 
\\
+(1-c_2)[A^{(n+1-q)}(*)_{n-1}(1-\delta_{q,1}) +A^{(n(1-\delta_{q,1})+2-q)}(*)_{n-1}(1-\delta_{q,2})]
\\
-(1-c_0)(1-c_2)[B^{(q-1)}_{n-2}(1-\delta_{q,n}) +B^{(q-2)}_{n-2}(1-\delta_{q,2})](1-\delta_{q,1}), \ q=1,...,n .
    \label{A^(q)_n}
\end{multline}
Eq.~(\ref{REA_(n)}) is the sum of Eqs.~(\ref{A^(q)_n}) over $q=1,2,...,n$ with a renamed $n \to n-1$.

Eq.~(\ref{REB_(n)}) is based on Lemma 2 applied to $B$ viewed as ${\tilde C}$ with two defects:
\begin{equation}
B_n = \tilde{C}_n + (1-c_0)A_{n-1} +(1-c_2)A(*)_{n-1} -(1-c_0)(1-c_2)B_{n-2} .
    \label{B_nViaTildeC_n}
\end{equation}
Namely, we use an analog of Eq.~(\ref{B_nViaTildeC_n}) for a matrix $B^{(q)}$ with a 1's-column defect:
\begin{multline}
B^{(q)}_n = \tilde{C}^{(q)}_n + (1-c_0)A^{(q)}_{n-1}(1-\delta_{q,n}) +(1-c_2)A^{(n+1-q)}(*)_{n-1}(1-\delta_{q,1})
\\
-(1-c_0)(1-c_2)B^{(q-1)}_{n-2}(1-\delta_{q,1})(1-\delta_{q,n}), \quad q=1,...,n .
    \label{B^(q)_n}
\end{multline}
Eq.~(\ref{REB_(n)}) is the sum of Eqs.~(\ref{B^(q)_n}) over $q=1,2,...,n$ with a renamed $n \to n-1$.

Eq.~(\ref{REC}) follows from Lemma 2 applied to $C^{(1)}$ viewed as $R$ with one defect:
\begin{equation}
C^{(1)}_n = R_n +(1-c_1)B_{n-1} = B_n +(1-c_1)B_{n-1} - (1-c_0)(1-c_2)B_{n-2} .
    \label{C^(1)_n}
\end{equation}

Eq.~(\ref{REA^(2)}) is Eq.~(\ref{A^(q)_n}) in the case $q=2$. When deriving the equations above, we plugged in some special permanents with a 1's column defect as follows
\begin{equation}
A^{(1)}_n =B^{(n)}_n = B^{(1)}(*)_n = A_n + (1-c_0)A_{n-1}, \quad A^{(n)}_n = A_{n-1} +A_{(n)}.
    \label{^(1)^(n)}
\end{equation}
This completes the proof of the Theorem.

{\bf Remark 3.3.} By means of changing the unknown variables or reducing or increasing their number, it is possible to rewrite the system of recurrence relations in other, more or less symmetric forms. For instance, it is straightforward to solve Eq.~(\ref{REB_(n)}) for the variable $B_{(n)}$ in terms of the basis permanents and exclude it from the system by plugging in it into the Eqs.~(\ref{REA_(n)})-(\ref{REC}) which, together with Eq.~(\ref{REA}) and the $(*)$-conjugated Eqs.~(\ref{*REA}), (\ref{*REA_(n)}), will constitute the six recurrence equations, each of the 4-th order, for the six unknown variables $A_n, A(*)_n, A_{(n)}, A(*)_{(n)}, B_n, C^{(1)}_n$. Overall, this system of linear homogeneous recurrence relations with variable coefficients is of the 24-th order. One can go further on and, similarly, exclude the variable $C^{(1)}_n$ by solving Eq.~(\ref{REA_(n)}) for $C^{(1)}_{n-1}$ and plugging in it into the remaining equations. A related analysis will be done elsewhere. Note that the actual overall order of this recurrence could be less than 24 only for some special cases or if there is some hidden symmetry or identity within this system of recurrence relations. In particular, for the problem of the 3-m{\'e}nage numbers (3-discordant permutations) corresponding to the case $c_0=c_1=c_2=0$, the system of recurrence relations can be reduced to just one recurrence Eq.~(\ref{straight3menageRE}) of the 8-th order for $A_n$ (see Remark 3.5 below).

{\bf Remark 3.4.} The system of recurrence relations (\ref{REA})--(\ref{REC}) is valid for $n \ge 7$ since the matrix size in the lowest-order permanent $A(*)_{n-4}$ entering Eqs.~(\ref{REA}), (\ref{REA_(n)}) should be equal or larger than the size of the diagonal band, $k=3$. The permanents of the order 6 and lower in the right hand side of the recurrence equations (that is, $A_3, A(*)_3, B_4, B_5$, etc.) should be computed directly via the definition of the permanent in Eq.~(\ref{per-def}). 
Direct numerical calculations confirm that the recurrence (\ref{REA})--(\ref{REC}) gives an easy and fast access to the correct result for the permanent of the circulant (\ref{C}) with arbitrary values of the entries $c_0, c_1, c_2$ for any, arbitrarily large matrix size $n$ in linear time.

{\bf Remark 3.5.} Famous 3-m{\'e}nage numbers \cite{Riordan1954,Yamamoto1956,Riordan1958,Whitehead1979,Canfield1987,OEIS,Zeilberger2014} $U^3_n$ counting the number of 3-discordant permutations $\sigma$ of $\{ 1,...,n\}$ such that $\sigma(j)$ is not congruent to any $j, j+1, j+2 \ (\textrm{mod} \, n)$, are equal to the permanent of the uniform circulant matrix $C$ with a band of three zero diagonals. Hence, they are given by a particular value, $U^3_n =C_n|_{c_0=c_1=c_2=0}$, of the permanent 
\begin{equation}
C_n = B_n -2A_{n-1} +B_{n-2}
    \label{U3=C_n}
\end{equation}
by means of the solution of the system of recurrence relations
\begin{equation}
A_n = A_{(n)} -C^{(1)}_{n-1} -A_{n-1} -A_{(n-2)} -2A_{n-2} -A_{n-3} +A_{n-4},
    \label{3menageREA}
\end{equation}
\begin{equation}
B_n = B_{(n)} -2A_{n-1} -2A_{n-2} +B_{n-2},
    \label{3menageREB}
\end{equation}
\begin{equation}
A_{(n)} = (n-2)C^{(1)}_{n-1} +A_{n-1} +2A_{(n-1)} +B_{(n-1)} +A_{(n-2)} -2B_{(n-2)} -A_{n-3},
    \label{3menageREA_(n)}
\end{equation}
\begin{equation}
B_{(n)} = (n-3)C^{(1)}_{n-1} +2A_{n-1} +2A_{(n-1)} -2B_{(n-2)} -2A_{n-3},
    \label{3menageREB_(n)}
\end{equation}
\begin{equation}
C^{(1)}_{n} = (n-3)C^{(1)}_{n-1} +B_{n-1} +2A_{(n-1)} -B_{(n-2)} -2A_{n-2} -2A_{n-3}
    \label{3menageREC}
\end{equation}
which are Eqs.~(\ref{REA})-(\ref{REC}) in the particular case $c_0=c_1=c_2=0$ when the $(*)$-conjugated counterparts are not needed since $A_n = A(*)_n$, $A_{(n)} = A(*)_{(n)}$.

We checked that the correct values (see the integer sequences A000183 and A001887 in \cite{OEIS}) and the known recurrence relations for the 3-m{\'e}nage numbers, or 3-discordant permutations,  \cite{Riordan1954,Yamamoto1956,Canfield1987,OEIS,Zeilberger2014} 
\begin{multline}
U^3_n = (-1)^n (4n+f_n) +\frac{n}{n-1}[(n+1)U^3_{n-1} +2(-1)^n f_{n-1}] -\frac{2n}{n-2}[(n-3)U^3_{n-2} 
\\
+(-1)^n f_{n-2}] + \frac{n}{n-3}[(n-5)U^3_{n-3} -2(-1)^n f_{n-3}] +\frac{n}{n-4}[U^3_{n-4} -(-1)^n f_{n-4}],
    \label{3menageRE} 
\end{multline}
(which is also valid for $n \ge 7$, see Example 4.7.17 in \cite{Stanley1-2012}) and for the straight 3-m{\'e}nage numbers \cite{Canfield1987,Flajolet2009} $V^3_n = A_n$ (which is valid for $n \ge 11$)
\begin{multline}
A_n = (n-1)A_{n-1} +(n+2)A_{n-2} -(3n-13)A_{n-3} - (2n-8)A_{n-4} 
\\
+(3n-15)A_{n-5} +(n-4)A_{n-6} -(n-7)A_{n-7} -A_{n-8}
    \label{straight3menageRE}
\end{multline}
follow from Eqs.~(\ref{U3=C_n})-(\ref{3menageREC}) derived above. (Eq.~(\ref{3menageRE}) includes the n-th Fibonacci number $F_n = F_{n-1} +F_{n-2}, F_0=0, F_1 =1$, via a function $f_n = F_{n+1} +F_{n-1} +2$.) In particular, in order to derive Eq.~(\ref{straight3menageRE}) one can, first, exclude $B_{(n)}$ by solving for it Eq.~(\ref{3menageREB}) and plugging in it into the remaining equations, then exclude $C^{(1)}_n$ by solving for it Eq.~(\ref{3menageREA_(n)}) and plugging in it into the remaining equations, then, in a similar way, exclude $A_{(n)}$, and, finally, exclude $B_n$.

\section{Conclusions} 

We present the exact solution for the circulant permanent via a finite system of the linear recurrence relations which provides a full access to a highly nontrivial analytic dependence of the permanent (whose entries are all nonzero) on $k=1, 2$ or $3$ independent parameters. This is especially interesting since computing the permanent in the case of 0-1 matrices with just three arbitrarily placed nonzero entries per row and column is as hard as in the general case \cite{Vazirani1988}. Exactly solvable models, like the ones discussed above, could play as important role in the understanding and theory of the matrix permanent and similar $\sharp P$-hard problems as the famous Onsager's and other exactly solvable models play in the theory of critical phenomena in phase transitions \cite{Kadanoff}. In other words, an attitude to the matrix permanent should be shifted from considering it just as a symbol of incomputability to employing it as a powerful tool for understanding and studying the $\sharp P$- and NP-hard problems and processes. 

The circulant permanent studied above is a multivariate polynomial of $k$ indeterminates $(c_0,..., c_{k-1})$ and has (say, in the case $k=3$) the following form  
\begin{equation}
\textrm{per}\,C = \sum_{j_0,j_1,j_2=0}^n P_{j_0,j_1,j_2} \ c_0^{j_0} c_1^{j_1} c_2^{j_2} ,
\label{polynomial}
\end{equation}
where summation over the indexes is subject to the constraint $\sum_{i=0}^{k-1} j_i \leq n$. The degree of such a permanental polynomial is equal to the matrix size $n$. The permanental polynomials constitute a key concept of universal polynomials in the algebraic complexity theory \cite{Burgisser1997} and their understanding is paramount.

The constant term of the permanental polynomial, $P_{j_0=0,...,j_{k-1}=0}$, considered as a function of the matrix size $n$ forms a famous integer sequence of the $k$-m{\'e}nage numbers, or $k$-discordant permutations \cite{Riordan1958,Whitehead1979,Canfield1987,OEIS,Zeilberger2014,Alekseyev2016}. The recurrence relations for the $k$-m{\'e}nage numbers are known in an explicit form only for the cases $k=1$ (derangements, or 1-m{\'e}nage numbers), $k=2$ (classical 2-m{\'e}nage numbers), and $k=3$ (3-m{\'e}nage numbers). The recurrence relations (\ref{REA})--(\ref{REC}) for the circulant permanent in the case $k=3$ provide a compact and explicit recurrence for the 3-m{\'e}nage numbers, Eqs.~(\ref{U3=C_n})-(\ref{3menageREC}) or (\ref{straight3menageRE}), if one simply sets all three circulant parameters to zero, $c_0=c_1=c_2=0$. Finding the recurrence for $k$-m{\'e}nage numbers in the case $k\ge4$ remains an open problem.

The finite recursion for the circulant permanents (presented above for the particular cases $k=1,2,3$) is the solution to a much more complex problem. It provides a powerful analytic and numerical tool for a detailed analysis of the entire permanental multivariate polynomial, not just its constant term that is the sequence of the $k$-m{\'e}nage numbers. The details of such an analysis will be given elsewhere. It would be interesting to generalize this approach to the case $k\ge4$ which is challenging. As the next step, we plan to write a separate paper on the aforementioned recursion in the case of the uniform circulant permanent with a band of $k=4$ any-value diagonals. 

For calculating the permanents, the method of recurrence relations for the permanents of matrices with defects, presented here, is superior to the rook and hit polynomial approach which dominated an extensive literature on the m{\'e}nage numbers (discordant permutations) in recent years \cite{Stanley1-2012,Riordan1958,Whitehead1979,Canfield1987,Bong1998,Zeilberger2014}. The point is that the rook and hit polynomials are somewhat artificial univariate polynomials associated with the permanents of just 0-1 matrices while the actual permanental polynomials, as in Eq.~(\ref{polynomial}), are multivariate ones and contain much more involved combinatorial and analytic information on permanents. Compared to the rook and hit polynomial approach, the permanent-based method is more algebraic rather than combinatorial in nature. 

The most important further step would be finding the asymptotics of the permanent for some classes of circulant constraint matrices of large size $n \gg 1$. The recurrence relations like the ones presented above allow one to apply well-developed asymptotic methods (see, for instance, \cite{Flajolet2009,WimpZeilberger1985,Zeilberger-MapleAsymptRecurrence,ApagoduZeilberger2006,ApagoduZeilberger-MultivariableAlgorithms} and references therein). A solution to this fundamental open problem on the permanent's asymptotics would be incredibly important for a unified analysis of a wide range of the nature's \text{\#}P-hard problems, including problems in the physics of many-body systems, critical phenomena, quantum computing, quantum field theory, theory of chaos, fractals, number theory, cryptography, etc. 

Finally, it is worth noting that the outlined above possibility to compute the permanent of the uniform circulant matrices with a band of $k$ diagonals in polynomial time does not contradict to the belief of an impossibility to compute the permanents in the general case of arbitrary matrices which is a \text{\#}P-hard problem according to the well-known Valiant's theorem \cite{Valiant1979} in the computational complexity theory. Although the permanents calculated above via the recurrence relations are very complex multivariant functions of the parameters $c_0,...,c_{k-1}$, a polynomial-time computability of such exactly solvable models is consistent with the principle of supremacy of quantum computing \cite{Kalai2016,ShchesnovichPRL2016,Harrow2017}. In fact, our ability to understand, describe and compute various complex processes of nature is increasing every time we bite off new computable or exactly solvable special cases from a general-case sea of NP- and \text{\#}P-hard problems.  

\section*{Acknowledgement}
We acknowledge the support by the Center of Excellence "Center of Photonics" funded by The Ministry of Science and Higher Education of the Russian Federation, contract № 075-15-2020-906.
     
\section*{Appendix: The $(*)$-conjugated recurrence equations (\ref{REA}) and (\ref{REA_(n)})}
\begin{equation}
A(*)_n = A(*)_{(n)} -(1-c_0)A(*)^{(2)}_{n-1} +(c_2+c_1-1)A(*)_{n-1} -(1-c_2)(1-c_1)A(*)_{n-2} ,
    \label{*REA}
\end{equation}
\begin{multline}
A(*)_{(n)} = (n-3)C^{(1)}_{n-1} +A(*)^{(2)}_{n-1} +A(*)_{n-1} +2(1-c_0)A_{(n-1)} +(1-c_1)B_{(n-1)}
\\
+(1-c_2)A(*)_{n-2} +(c_1+c_0-2)A_{n-2} -c_0(1-c_0)A_{(n-2)} -2(1-c_0)(1-c_2)B_{(n-2)}
\\
-(2+c_2-c_1-c_0)(1-c_0)A_{n-3} +(1-c_2)(1-c_0)^2 A_{n-4} ,
\label{*REA_(n)}
\end{multline}
where
\begin{multline}
A(*)^{(2)}_{n-1} = C^{(1)}_{n-1} +(1-c_1)A_{n-2} +(1-c_0)A_{(n-2)} 
\\
+ (1+c_2-c_1)(1-c_0)A_{n-3} -(1-c_2)(1-c_0)^2 A_{n-4} .
    \label{*REA^(2)}
\end{multline}


\begin{thebibliography}{99}

\bibitem{Entropy2020} V.V. Kocharovsky, Vl.V. Kocharovsky, S.V. Tarasov, Unification of the Nature’s Complexities via a Matrix Permanent -- Critical Phenomena, Fractals, Quantum Computing, \text{\#}P-Complexity, Entropy \textbf{22} (2020) 322. 

\bibitem{Caianiello1973} E.R. Caianiello, Combinatorics and renormalization in quantum field theory, in: Frontiers in Physics. W. A. Benjamin Inc, Reading, London, 1973.

\bibitem{Wosiek1997} J. Wosiek, A simple formula for Bose-Einstein corrections, Phys. Lett. B \textbf{399} (1997) 130-134.

\bibitem{Scheel2004} S. Scheel, Permanents in linear optical networks, arXiv:quant-ph/0406127v1 18Jul2004.

\bibitem{Aaronson2011} S. Aaronson, A linear-optical proof that the permanent
is $\sharp$P-hard, Proc. R. Soc. A \textbf{467} (2011) 3393--3405.

\bibitem{Kalai2016} G. Kalai, The quantum computer puzzle (expanded version), arXiv:quant-ph/1605.00992 3May2016.

\bibitem{ShchesnovichPRL2016} V.S. Shchesnovich, Universality of Generalized Bunching and Efficient Assessment of Boson Sampling, Phys. Rev. Lett. \textbf{116} (2016) 123601.

\bibitem{Drummond2018} B. Opanchuk, L. Rosales-Zárate, M.D. Reid, P.D. Drummond, Simulating and assessing boson sampling experiments with phase-space representations, Phys. Rev. A \textbf{97} (2018) 042304.  

\bibitem{Kadanoff} L.P. Kadanoff, Statistical Physics: Statics, Dynamics and Renormalization, World Scientific, Singapore, 2000.

\bibitem{arXivIsing2016} V.V. Kocharovsky, Vl.V. Kocharovsky, Exact general solution to the three-dimensional Ising model and a self-consistency equation for the nearest-neighbors' correlations, arXiv:cond-mat.stat-mech/1510.07327v3 24Mar2016.

\bibitem{PhysicaScripta2015} V.V. Kocharovsky, Vl.V. Kocharovsky, Microscopic theory of phase transitions in a critical region, Physica Scripta \textbf{90} (2015) 108002.

\bibitem{PLA2015Ising} V.V. Kocharovsky, Vl.V. Kocharovsky, Towards an exact solution for the three-dimensional Ising model: A method of the recurrence equations for partial contractions, Phys. Lett. A \textbf{379} (2015) 2520--2523.

\bibitem{Minc1978} H. Minc, Permanents, Encyclopedia of Math. and its Appl. v. 6, Addison - Wesley, Reading, 1978.

\bibitem{Minc1987} H. Minc, Theory of Permanents 1982-1985, Linear and  Multilinear Algebra \textbf{21} (1987) 109--148.

\bibitem{Bapat2007} R.B. Bapat, Recent developments and open problems in the theory of permanents, Math. Student \textbf{76} (2007) 55--69.

\bibitem{Brualdi2008} R.A. Brualdi, D. Cvetkovic, A combinatorial approach to matrix theory and its applications, CRC Press, Boca Raton, 2008.

\bibitem{Stanley1-2012} R.P. Stanley, Enumerative Combinatorics, v.1, Cambridge Univ. Press, Cambridge, 2012.

\bibitem{Barvinok2016} A. Barvinok, Combinatorics and Complexity of Partition Functions, Algorithms and Combinatorics 30, Springer International Publishing AG, 2016.

\bibitem{Ryser} H.J. Ryser, Combinatorial mathematics, The Carus Mathematical Monographs, No. 14, The Mathematical Association of America, 1963.

\bibitem{Valiant1979}  L.G. Valiant, The complexity of computing the permanent, Theor. Comput. Sci. \textbf{8} (1979) 189--201.

\bibitem{Jerrum2004} M. Jerrum, A. Sinclair, E. Vigoda, A polynomial-time approximation algorithm for the permanent of a matrix with nonnegative entries, Journal of the ACM \textbf{51} (2004), no.
4, 671–-697.

\bibitem{Jerrum2015} L.A. Goldberg, M. Jerrum, A complexity classification of spin systems with an external field, Proc. Nat. Acad. Sci. \textbf{112} (2015) 13161--13166.

\bibitem{Diaconis1999} P. Diaconis, R. Graham, S.P. Holmes, Statistical problems involving permutations with restricted positions, 1999, available at http://www-stat.stanford.edu/ cgates/PERSI/papers/perm8.ps.

\bibitem{GrahamLehmer1976} R.L. Graham, D.H. Lehmer, On the permanent of Schur's matrix, J. Austral. Math. Soc. A \textbf{21} (1976) 487--497.

\bibitem{Burgisser1997} P. Bürgisser, M. Clausen, M.A. Shokrollahi, Algebraic Complexity Theory. With the collaboration of Thomas Lickteig, Grundlehren der MathematischenWissenschaften
[Fundamental Principles of Mathematical Sciences], 315, Springer-Verlag, Berlin, 1997.

\bibitem{Gurvits2002} L. Gurvits, A. Samorodnitsky, A Deterministic Algorithm for Approximating the Mixed Discriminant and Mixed Volume, and a Combinatorial Corollary, Discrete Comput. Geom. \textbf{27} (2002) 531--550. 

\bibitem{Gurvits2017} N. Anari, L. Gurvits, S.O. Gharan, A. Saberi, Simply Exponential Approximation of the Permanent of Positive Semidefinite Matrices, in: Foundations of Computer Science (FOCS), 2017 IEEE 58th Annual Symposium, 2017, pp. 914--925; DOI 10.1109/FOCS.2017.89 

\bibitem{Fyodorov2006} Y.V. Fyodorov, On permanental polynomials of certain random matrices, Intern. Math. Research Notices Article ID 61570 (2006).

\bibitem{Schwartz2009} M. Schwartz, Efficiently computing the permanent and Hafnian of some banded Toeplitz matrices, Linear Algebra Appl. \textbf{430} (2009) 1364--1374.

\bibitem{FiedlerHall2012} M. Fiedler, F.J. Hall, A note on permanents and generalized complementary basic matrices, Linear Algebra Appl. \textbf{436} (2012) 3553--3561.

\bibitem{Shchesnovich2016} V.S. Shchesnovich, The permanent-on-top conjecture is false, Linear Algebra Appl. \textbf{490} (2016) 196--201.

\bibitem{Cifuentes2016} D. Cifuentes, P.A. Parrilo, An efficient tree decomposition method for permanents and mixed discriminants, Linear Algebra Appl. \textbf{493} (2016) 45--81.

\bibitem{Glynn2010} D.G. Glynn, The permanent of a square matrix, Eur. J. Combinatorics \textbf{31} (2010) 1887--1891.

\bibitem{Landsberg-JPAA2017} C. Ikenmeyera, J.M. Landsberg, On the complexity of the permanent in various computational models, J. Pure Appl.Algebra \textbf{221} (2017) 2911--2927.

\bibitem{King1969} B.W. King, F.D. Parker, A Fibonacci matrix and the permanent function, Fibonacci Quart. \textbf{7} (1969) 539544.

\bibitem{Minc1987LAA} H. Minc, Permanental compounds and permanents of (0,1) circulants. Linear Algebra Appl. \textbf{86} (1987) 11--42.

\bibitem{Codenotti1999} A. Bernasconi, B. Codenotti, V. Crespi, G. Resta, How fast can one compute
the permanent of circulant matrices? Linear Algebra Appl. \textbf{292} (1999) 15--37.

\bibitem{Thomas2004} H. Thomas, The number of terms in the permanent and the determinant of a generic circulant matrix, J. Algebraic Combinatorics \textbf{20} (2004) 55-60.

\bibitem{LAA2017} V.V. Kocharovsky, Vl.V. Kocharovsky, On the permanents of circulant and degenerate Schur matrices, Linear Algebra Appl. \textbf{519} (2017) 366--381. 

\bibitem{Poi-JPAA2019} P. De Poi, E. Mezzetti, M. Michałek, R.M. Mir{\'o}-Roige, E. Nevo, Circulant matrices and Galois-Togliatti systems, J. Pure Appl. Algebra \textbf{224} (2019) 106404.

\bibitem{Codenotti2001} B. Codenotti, G. Resta, On the permanent of certain circulant matrices. In: H. Crapo et al. (eds.), Algebraic Combinatorics and Computer Science. Springer-Verlag Italia, Milano, 2001. 

\bibitem{Fonseca2010} C.M. da Fonseca, The $\mu$-permanent of a tridiagonal matrix, orthogonal polynomials, and chain sequences, Linear Alg. Appl. \textbf{432} (2010) 1258--1266.

\bibitem{Temme2012} K. Temme, P. Wocjan, Efficient Computation of the Permanent of Block Factorizable Matrices, arXiv:1208.6589v1 [cs.DM] 31Aug2012. 

\bibitem{Butera2015} P. Butera, M. Pernici, Sums of permanental minors using Grassmann algebra, Intern. J. Graph Theory Appl. \textbf{1} (2015) 83--96.

\bibitem{Servedio2005} R.A. Servedio, A. Wan, Computing sparse permanents faster, Information Processing Letters \textbf{96} (2005) 89--92.

\bibitem{Goldenfeld} N. Goldenfeld, Lectures on Phase Transitions and Renormalization Group, Reading, MA: Addison-Wesley, 1992.

\bibitem{Riordan1958} J. Riordan, Introduction to Combinatorial Analysis, John Wiley, 1958. Reprinted by Dover, 2002.

\bibitem{Whitehead1979} E.G. Whitehead, Jr., Four-discordant permutations, J. Austral. Math. Soc. (Series A) \textbf{28} (1979) 369--377.

\bibitem{Canfield1987} E.R. Canfield, N.C. Wormald, M{\'e}nage numbers, bijections and P-recursiveness, Discrete Math. \textbf{63} (1987) 117--129.

\bibitem{Bong1998} Nguyen‐Huu Bong, Lucas numbers and the m{\'e}nage problem, Internat. J. of Math. Education in Science and Technology \textbf{29} (1998) 647--661.

\bibitem{Flajolet2009} P. Flajolet, R. Sedgewick, Analytic Combinatorics, Cambridge Univ. Press, 2009.  

\bibitem{OEIS} The Online Encyclopedia of Integer Sequences, internet resource, https://oeis.org/.

\bibitem{Zeilberger2014} D. Zeilberger, Automatic Enumeration of Generalized M{\'e}nage Numbers, arXiv:math.CO/1401.1089v1 6Jan2014.

\bibitem{Alekseyev2016} M.A. Alekseyev, Weighted de Bruijn Graphs for the M{\'e}nage Problem and Its
Generalizations, arXiv:math.CO/1510.07927v4 9Aug2016.

\bibitem{Muir1882} T. Muir, Additional note on a problem of arrangement, Proceedings of the Royal Society of Edinburgh \textbf{11} (1882) 187--190.

\bibitem{KaplanskyRiordan1946} I. Kaplansky, J. Riordan, The probleme des menages, Scripta Math. \textbf{12} (1946) 113--124.

\bibitem{Touchard1953} J. Touchard, Permutations discordant with two given permutations, Scripta Math. \textbf{19} (1953) 109--119.

\bibitem{Riordan1954} J. Riordan, Discodant permutations, Scripta Math. \textbf{20} (1954) 14--23.

\bibitem{Yamamoto1956} K. Yamamoto, Structure polynomial of Latin rectangles and its application to a combinatorial problem, Memoirs of the Faculty of Science, Kyusyu University, Series A \textbf{10} (1956) 1--13.

\bibitem{Vazirani1988} P. Dagum, M. Luby, M. Mihail, U. Vazirani, Polytopes, permanents, and graphs with large factors, in: Proceedings of the 29th IEEE Symposium on Foundations of Computer Science. Institute of Electrical and Electronics Engineers. Washington, DC, 1988,
412--421.

\bibitem{WimpZeilberger1985} J. Wimp, D. Zeilberger, Resurrecting the Asymptotics of Linear Recurrences, J. Math. Anal. Appl. \textbf{111} (1985) 162--176.

\bibitem{Zeilberger-MapleAsymptRecurrence} D. Zeilberger, AsyRec, a Maple package that computes the asymptotics for solutions of linear recurrence equations with polynomial coefficients; available at $http://www.math.rutgers.edu/^{\sim}zeilberg/tokhniot/AsyRec$.

\bibitem{ApagoduZeilberger2006} M. Apagodu, D. Zeilberger, Multi-variable Zeilberger and Almkvist-Zeilberger algorithms and the sharpening of Wilf-Zeilberger theory, Adv. Appl. Math. \textbf{37} (2006) 139--152. 

\bibitem{ApagoduZeilberger-MultivariableAlgorithms} M. Apagodu,  D. Zeilberger, Multi-variable Zeilberger and Almkvist-Zeilberger algorithms and the Sharpening of Wilf-Zeilberger Theory, software package, available at {\it http://www.math.} $rutgers.edu/^{\sim}zeilberg/mamarim/mamarimhtml/multiZ.html$.

\bibitem{Harrow2017} A.W. Harrow, A. Montanaro, Quantum computational supremacy, Nature \textbf{549} (2017) 203.

\end{thebibliography}
\end{document}